\documentclass[letterpaper]{ptephy_v1}

\usepackage{boites}
\usepackage{graphicx}
\usepackage{bm}
\usepackage{amsmath,amssymb,amsfonts}

\newcommand{\nc}{\newcommand}		
\nc{\vc}[1]	{\mbox{\boldmath $#1$}}	
\nc{\bra}       {\langle}               
\nc{\ket}       {\rangle}               
\nc{\bras}[1]   {\langle#1|}            
\nc{\kets}[1]   {|#1\rangle}            
\nc{\del}       {\partial}              
\nc{\beq}       {\begin{eqnarray}}
\nc{\eeq}       {\end{eqnarray}}
\nc{\AMD}       {{\rm AMD}}
\nc{\TOAMD}     {{\rm TOAMD}}
\newcommand{\lw}[1]{\smash{\lower1.75ex\hbox{#1}}}

\usepackage{color}  
\nc{\red}[1]    {\textcolor{red}{#1}}  

\begin{document}

\title{Hybridization of tensor-optimized and high-momentum antisymmetrized molecular dynamics for light nuclei with bare interaction}

\author{
\name{Mengjiao Lyu}{1}, \name{Masahiro Isaka}{1}, \name{Takayuki Myo}{1,2}, \name{Hiroshi Toki}{1},  \name{Kiyomi Ikeda}{3}, \name{Hisashi Horiuchi}{1}, \name{Tadahiro Suhara}{4},   and \name{Taiichi Yamada}{5}
}

\address{
\affil{1}{1. Research Center for Nuclear Physics (RCNP), Osaka University, Ibaraki, Osaka 567-0047, Japan\\}
\affil{2}{2. General Education, Faculty of Engineering, Osaka Institute of Technology, Osaka, Osaka 535-8585, Japan\\}
\affil{3}{3. RIKEN Nishina Center, Wako, Saitama 351-0198, Japan\\}
\affil{4}{4. Matsue College of Technology, Matsue 690-8518, Japan\\}
\affil{5}{5. Laboratory of Physics, Kanto Gakuin University, Yokohama 236-8501, Japan\\}
\email{mengjiao@rcnp.osaka-u.ac.jp}
\email{takayuki.myo@oit.ac.jp}}

\begin{abstract}%
Many-body correlations play an essential role in the {\it ab initio} description of nuclei with nuclear bare interactions. We propose a new framework to describe light nuclei by the hybridization of the tensor-optimized antisymmetrized molecular dynamics (TOAMD) and the high-momentum AMD (HM-AMD), which we call ``HM-TOAMD". In this framework, we describe the many-body correlations in terms of not only the correlation functions in TOAMD, but also the high-momentum pairs in the AMD wave function. With the bare nucleon-nucleon interaction AV8$^\prime$, we sufficiently reproduce the energy and radius of the ${^3}$H nucleus in HM-TOAMD. The effects of tensor force and short-range repulsion in the bare interaction are nicely described in this new framework. We also discuss the convergence in calculation and flexibility of the model space for this new method.
\end{abstract}

\subjectindex{ D10,  D11}


\maketitle

\noindent{\mbox{1.\,\, {\it Introduction}}}~
In {\it ab initio} description of nuclei, nucleon-nucleon (NN) correlations are mainly induced by the strong short-range repulsion and strong tensor force in bare NN interaction \cite{wiringa95,myo15}. Many-body effects of these NN correlations play essential roles for the accurate description of the nuclear system.

In our recent works, two different approaches have been established for the description of the NN correlations in {\it ab initio} calculations of {\it s}-shell nuclei with bare interactions \cite{myo15,myo17a,myo17b,myo17c,myo17d}. These studies are based on the framework of antisymmetrized molecular dynamics (AMD) which has been very successful in the investigations of nuclear clustering systems \cite{kanada03,kanada12}. One of our approaches is the tensor-optimized antisymmetrized molecular dynamics (TOAMD), which successfully describes the {\it s}-shell nuclei with bare interaction \cite{myo15,myo17a,myo17b,myo17c,myo17d}. In the TOAMD wave function, short-range and tensor correlations are described by using the correlation functions condescending to the short-range repulsion and tensor force.
By including up to the double products of the correlation functions, the TOAMD results nicely reproduce those obtained from the Green's function Monte Carlo (GFMC) method. As a next step, we will apply this TOAMD framework to $p$-shell nuclei, but massive analytical derivations and computation time are necessary to handle. Another very recent progress in our group is the superposition of the AMD basis states including the nucleon pairs with high momentum (HM) components, which we name HM-AMD. In the study of $s$-shell nuclei in HM-AMD, it is found that the superposition of the single nucleon-pair with various high momentum components provides the equivalent effect of the two-particle two-hole excitation of the tensor correlation \cite{myo17e,itagaki17}. 

In the present work, we propose a new framework of the hybridization of the tensor-optimized and the high-momentum AMD, namely ``HM-TOAMD". One important advantage of this combination is that this method can describe more effectively the many-body correlations in light nuclei compared to TOAMD. Especially, for {\it ab initio} calculation of {\it p}-shell and heavier nuclei, many kinds of correlation terms have to be included in the wave function, while the high-momentum pairs in HM-TOAMD are much easier to handle in the formulation part as compared to the naive extension of the TOAMD method. In addition, the flexibility of this method provides the possibility to select only essential basis states to express the wave function; the important pair correlations can be described from the selection of the appropriate high-momentum pairs and the additional multi-nucleon correlations can be expressed using the correlation functions in the TOAMD part.  With this simplicity and flexibility, the new HM-TOAMD is promising for future studies not only of $p$-shell and cluster nuclei, but also for the treatment of three-nucleon interactions.

In the following paragraphs, the formulation of HM-TOAMD are provided. After that, the effects of short-range and tensor correlations are clearly demonstrated by showing the results of energy curves of $^{3}$H with respect to different imaginary shifts of a high-momentum pair. The energies and radius of the ${^3}$H nucleus are reproduced with HM-TOAMD using the bare AV8$^\prime$ interaction. In HM-TOAMD, there are two kinds of variational functions; one is high-momentum pair and another is correlation functions. The convergence with respect to these variational functions and flexibility of the model space in this new method are also discussed. The final paragraph contains the conclusion.

\vspace*{0.2cm}
\noindent{\mbox{2.\,\, {\it Formulation}}}~
We start writing the AMD basis of $A$ nucleons to be constructed with the antisymmetrization expressed by Slater determinant ($\det$) of single nucleon states \cite{myo15}:
\begin{equation}\label{amd}
|\Phi_{\rm AMD}\rangle=
\det \{|\phi_{1}(\vec r_1) \cdots \phi_{A}(\vec r_A)\rangle\},
\end{equation}
where the single-nucleon states $\phi(\vec{r})$ with spin $\sigma$ and isospin $\tau$ have the form of the Gaussian wave packet with the centroid $\vec Z$ and $\chi_{\tau,\sigma}$ for spin-isospin components.
\begin{eqnarray}\label{single-state}
\phi(\vec r)\propto
e^{-\nu(\vec r-\vec Z)^2 } \chi_{\tau,\sigma}.
\end{eqnarray}

We explain first the AMD basis states with high-momentum components in the HM-AMD method \cite{myo17e}. We take the case of $^{3}$H in this paper and consider three kinds of basis states for HM-AMD, which are denoted by the following expressions:
\begin{equation}\label{FH}
1: \vec D_{p\uparrow,n\uparrow  },\quad
2: \vec D_{p\uparrow,n\downarrow},\quad
3: \vec D_{n\uparrow,n\downarrow  }.\quad
\end{equation}
Here, we drop the 4-th basis state $\vec D_{n\uparrow,n\uparrow}$ due to the assumption that the spatial wave function is symmetric and the wave function should be antisymmetrized in the spin-isospin space.
In the above symbols, $\vec D$ is the vector of the imaginary shift for the Gaussian centroid defined as $\vec{Z}=\vec{R}+i\vec{D}$, in which the vector $\vec R$ generally represents the spatial position of the centroid.  In the HM-AMD, each $\vec D$ should be assigned to two nucleon pair with the same magnitude $D=|\vec D|$ and opposite direction \cite{myo17e,itagaki17}. 
We take various values for the imaginary shift $D$ for each pair, which are called as ``high-momentum pairs", and superpose these AMD basis states in HM-AMD.  When we take the basis number $n_{\rm D}$ for each sign of $\vec D$ as $\pm\vec D$, the total basis number for high-momentum configurations is $6\times n_{\rm D}$.
We write explicitly the case 1 of Eq.~(\ref{FH}) as
\begin{eqnarray}\label{f-hm}
\phi_{p\uparrow} (\vec r_1 ) \phi_{n\uparrow} (\vec r_2 )\propto 
  e^{-\nu(\vec r_1-\vec R_1\pm i \vec D{_{p\uparrow n\uparrow}})^2 } \chi_{p\uparrow}\,
  e^{-\nu(\vec r_2-\vec R_2\mp i \vec D{_{p\uparrow n\uparrow}})^2 } \chi_{n\uparrow}.
\end{eqnarray}
Here $\phi_{p\uparrow} (\vec r_1 )$ and $\phi_{n\uparrow} (\vec r_2 )$ are the single nucleon states of proton and neutron with imaginary shift $\vec D{_{p\uparrow n\uparrow}}$. Eq.~(\ref{f-hm}) can be rewritten in terms of the relative $\vec r$ and center-of-mass $\vec R$  coordinates as 
\begin{equation}
\phi_{p\uparrow} (\vec r_1 ) \phi_{n\uparrow} (\vec r_2 )\propto 
e^{-\nu/2(\vec r-\vec Z_r)^2}
e^{-2\nu(\vec R-\vec Z_R)^2}\chi_{p\uparrow}\chi_{n\uparrow},\\
\end{equation}
where $\vec r= \vec{r}_1 - \vec{r}_2$, $\vec R=(\vec {r}_1 + \vec {r}_2)/2$, and the centroid parameters are
\begin{equation}
\vec Z_r= \vec R_1 - \vec R_2 \pm 2i\vec D{_{p\uparrow n\uparrow}}, \,\,\,\,
\vec Z_R=(\vec R_1 + \vec R_2)/2.
\end{equation}
Clearly, it is found that the imaginary shifts $\vec D{_{p\uparrow n\uparrow}}$ only change the relative parameter $\vec{Z}_r$ between two nucleons of the high momentum pair.  In this way, we introduce correlations among two nucleons in the HM-AMD \cite{myo17e}. For the ground states of $s$-shell nuclei, the optimized value for the spatial centroid parameters are $\vec R_i=\vec{0}$.

We discuss now another way to include correlations among a nucleon pair, which is the TOAMD method.  The TOAMD wave function in its lowest order (Single-$F$ TOAMD) is given as \cite{myo15}:
\begin{equation}\label{toamd}
(1+F_S+F_D)\times |\Psi_{\rm AMD}\rangle.
\end{equation}
The two-body correlation function $F_S$ for the short-range correlation and $F_D$ for the tensor correlation are expressed as the superposition of Gaussian functions. This form of TOAMD can be extended by increasing the power of the correlation functions successively. This extension improves the variational accuracy of TOAMD, but increases the computational costs, and the formulation and programing costs tremendously as the power of the correlation increases.  On the other hand, in HM-AMD, we superpose various directions and magnitudes of the relative momentum components for any nucleons pairs in the AMD basis states.  It is much easier to increase the number of correlations in the wave function using the technologies developed for the AMD framework.

We propose here to combine the two methods: TOAMD and HM-AMD.  In this framework named as HM-TOAMD, we write the total wave function as:
\begin{equation}\label{hm-toamd}
|\Psi_{\rm HM-TOAMD} \rangle=
\sum_\alpha 
\left( C_\alpha
+\sum_{t=0}^1 \sum_{s=0}^1 C^{t,s}_{S,\alpha}F^{t,s}_{S,\alpha}
+\sum_{t=0}^1              C^t_{D,\alpha} F^t_{D,\alpha}\right) 
\hat{P}^J_{MK}|\Psi_{\rm HM-AMD,\alpha}\rangle.
\end{equation}
Eq.~(\ref{hm-toamd}) is given in the linear combination form of the basis states of HM-TOAMD. $\Psi_{\rm HM-AMD}$ is the HM-AMD wave function and the label $\alpha$ is to distinguish the basis states including both the high-momentum components in $\Psi_{\rm HM-AMD}$ and the correlation functions. The $\hat{P}^{J}_{MK}$ is the projection operator to total angular momentum state $J$ \cite{schuck80}. The labels $t$ and $s$ are the isospin and spin values of the correlated two nucleons. The $C_{\alpha}$, $C^{t,s}_{S,\alpha}$ and $C^{t}_{D,\alpha}$ are the superposition coefficients. The tensor part $F^t_{D,\alpha}$ contains two-channel tensor-operator terms and the central part $F^{t,s}_{S,\alpha}$ contains four-channel short-range central terms. The radial forms of the correlation function are the Gaussian functions with spin-isospin operators \cite{myo17a}:
\begin{eqnarray}
\label{f-toamd}
&&F^t_{D,\alpha}= \sum_{i<j}^A \exp(-a^t_{D,\alpha} r_{ij}^2) O_{ij}^t r_{ij}^{2} S_{12}(\hat{r}_{ij}),\\
&&F^{t,s}_{S,\alpha}= \sum_{i<j}^A \exp(-a^{t,s}_{S,\alpha} r_{ij}^2) O_{ij}^t O_{ij}^s.
\end{eqnarray}
Here, $\vec r_{ij}=\vec r_i-\vec r_j$, $O_{ij}^t=(\vec \tau_i \cdot
\vec \tau_j)^t$ and $O_{ij}^s=(\vec \sigma_i \cdot
\vec \sigma_j)^s$.
In each channel, we superpose the Gaussian functions with the common basis number $n_{\rm G}$. Hence, the total number of basis states coming from the bracket term in Eq.~(\ref{toamd}) as TOAMD is $1+6\times n_{\rm G}$.
In total, the total basis number for HM-TOAMD is $(1+6\times n_{\rm G})\times 6 \times n_{\rm D}$ in the GCM-type calculation in Eq.~(\ref{hm-toamd}).

Finally, we solve the eigenvalue problem in the following equation and determine $\tilde C_\alpha=\{C_\alpha,C_\alpha^t, C_\alpha^{t,s}\}$ in the linear combination form.
\begin{eqnarray}\label{hw}
&&|\Psi_{\rm HM-TOAMD} \rangle =\sum_\alpha 
\tilde C_\alpha |\Psi_{\rm HM-TOAMD,\alpha} \rangle,
\\
&&\sum_{\alpha,\beta}\left( H_{\alpha \beta}-EN_{\alpha \beta}\right) \tilde C_\beta =0,
\\
&&\langle \Psi_{\rm HM-TOAMD,\alpha} |H| \Psi_{\rm HM-TOAMD,\beta}\rangle =H_{\alpha \beta},
\\
&&\langle \Psi_{\rm HM-TOAMD,\alpha} |   \Psi_{\rm HM-TOAMD,\beta}\rangle =N_{\alpha \beta}.
\end{eqnarray}
{ The bases $|\Psi_{\rm HM-TOAMD, \alpha} \rangle$ are carefully chosen to reflect the physical properties of NN correlations. For the Gaussian range parameters $a$ in Eqs.~(9) and (10), a series of bases ranging from 0.01 fm${}^{-2}$ to 5 fm${}^{-2}$ are adopted and optimized for each channel by Single-F TOAMD calculation. For the imaginary shifts, absolute values $\vec D$ ranging from 1 fm to 12 fm are adopted for each kind of high-momentum pair, with corresponding momentum $\left<k_z\right>=2\nu\cdot \rm{Im}(D_z)$ ranging from 0.28 fm$^{-1}$  to 2.8 fm$^{-1}$ using $\nu$=0.14 fm$^{-2}$. }

\begin{figure}[t]
\centering
\includegraphics[width=12.0cm,clip]{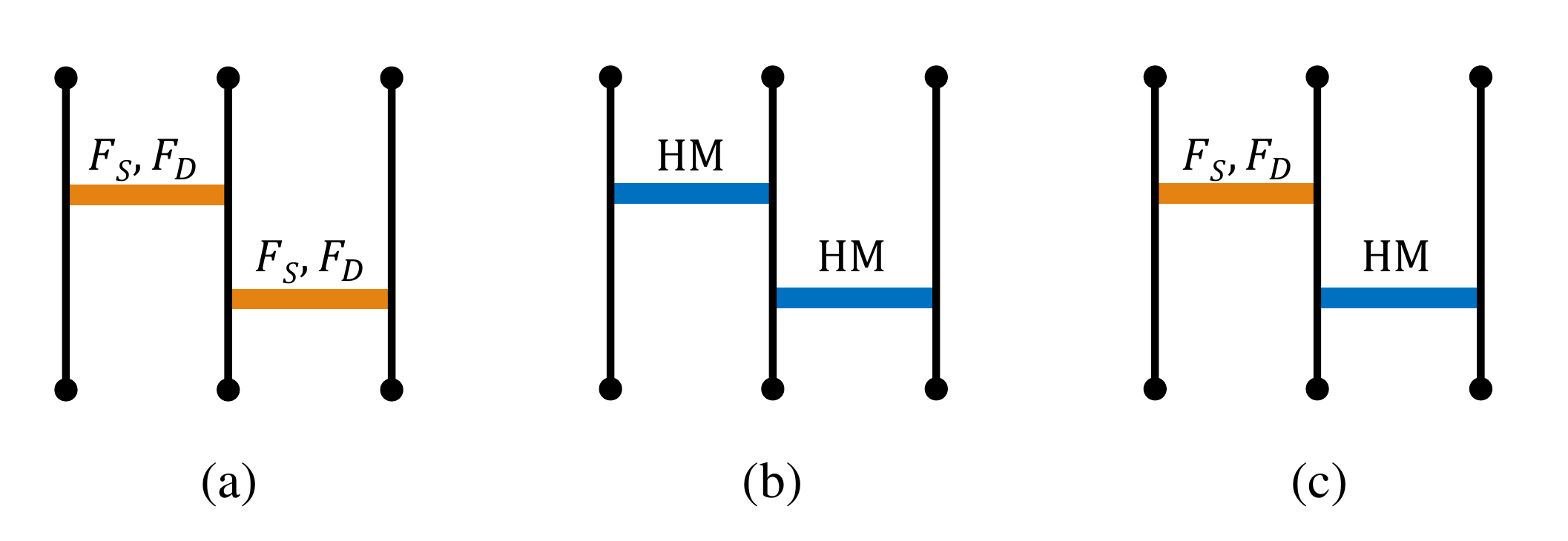}
\caption{Diagrams of the three-body correlations in double products of correlation functions for ${^3}$H: (a) TOAMD, (b) HM-AMD, (c) HM-TOAMD. ``$F_S, F_D$" denote the NN correlations described by correlation functions in Eqs.~(7) and (8) and ``HM" denotes the NN correlations described by high-momentum pairs in Eqs.~(\ref{FH}) and (\ref{f-hm}).}
\label{fig:double-f}
\end{figure}

We discuss the difference between the new HM-TOAMD approach and our previous TOAMD and HM-AMD by showing the three-body correlations in case of ${^3}$H in Figure ~\ref{fig:double-f}. In this figure, the case (a) shows the ``Double-$F$ TOAMD" approach \cite{myo17a,myo17b,myo17c,myo17d}, which includes the double products of two kinds of the NN correlations described by functions $F_{D}$ and $F_{S}$ as in Eqs.~(9) and (10). The case (b) shows the pure HM-AMD approach \cite{myo17f,isaka17}, in which all the correlations are described by using the double high-momentum pairs as in Eqs.~(\ref{FH}) and (\ref{f-hm}). In our new hybrid description HM-TOAMD, one of the correlation is described by the correlation functions $F_{D}$ and $F_{S}$ and another one is described by the high-momentum pairs, as shown in the case (c).

\begin{figure}[t]
\centering
\includegraphics[width=11.0cm,clip]{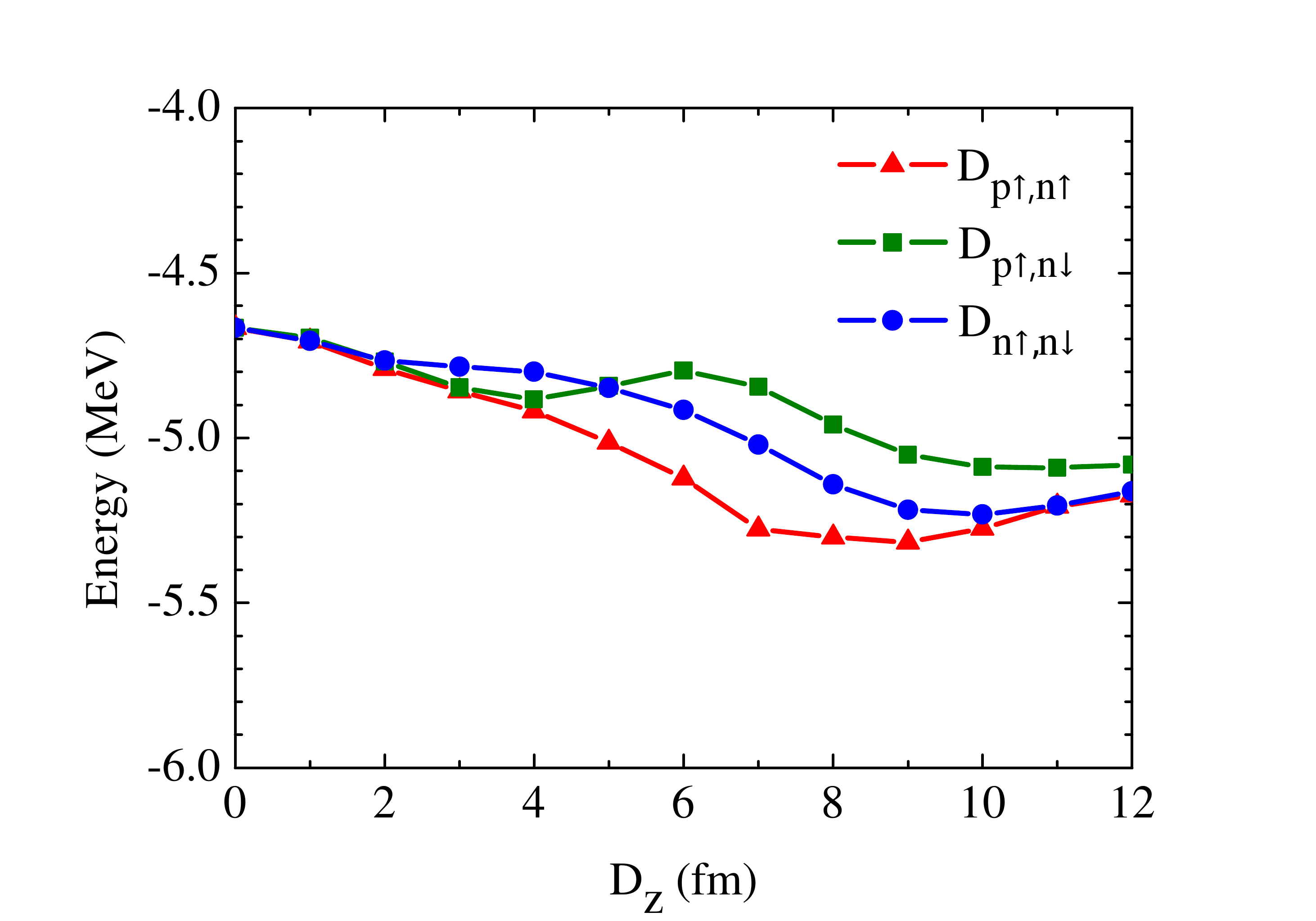}
\caption{Energy curves of the ${^3}$H nucleus calculated with HM-TOAMD and the bare interaction AV8$^\prime$ with respect to the shift $D_z$ in the $z$-direction when only a single high-momentum pair is included as $n_D=1$. The symbols $D_{p\uparrow,n\uparrow}$, $D_{p\uparrow,n\downarrow}$ and $D_{n\uparrow,n\downarrow}$ denote different kinds of high-momentum pairs with positive shift $D$ as in Eq.~(\ref{f-hm}). We use $n_{\rm G}=3$  and $\nu=0.14$ fm$^{-2}$.}
\label{fig:e-dz}
\end{figure}

\vspace*{0.2cm}
\noindent{\mbox{3.\,\, {\it Results}}}~
We discuss first the effect of the imaginary shift $\vec D$ on the total energy for $^3$H. 
In our previous study of HM-AMD, we have shown that the imaginary shift $|\vec D|=5$ fm {($k_z=$2.5 fm$^{-1}$)} is energetically favored by the tensor interaction \cite{myo17e}. 
In our new hybrid approach, we can discuss the effects of $\vec D$ in more accurate situation, where we include the single correlation function in terms of TOAMD and the second correlation in terms of the high-momentum pair as shown in Figure \,\ref{fig:double-f} (c).  Figure \,\ref{fig:e-dz} shows the energy curves of ${^3}$H for three kinds of high-momentum pairs as functions of $D_z=|\vec D|$ in the $z$-direction which is parallel to the spin direction of nucleons, with $n_{\rm D}$=1, where we employ the number of Gaussian $n_{\rm G}=3$ in $F_D$ and $F_{S}$ of the TOAMD part. We use the AV8$^\prime$ bare interaction. In this figure, we see clearly two minima around $D_z=4$ fm {($k_z=$1.12 fm$^{-1}$)} and $D_z=10$ fm {($k_z=$2.8 fm$^{-1}$)} for the $(p\uparrow, n\downarrow)$ high-momentum pair, which are caused by the effects of the tensor interaction and the short-range repulsion in the AV8$^\prime$ interaction, respectively. For the $(p\uparrow, n\uparrow)$ high-momentum pair, we obtain the minimum around $D_z=8$ fm corresponding to the short-range effect.  For these results, we should keep in mind that both the strong short-range effect and the tensor correlation have been already included in the correlation function $F_S$ and $F_D$ of TOAMD, respectively.  For the $(n\uparrow, n\downarrow)$ high-momentum pair, the tensor correlation is weak and then we can confirm only the  effect of short-range correlation with the energy minimum around $D_z=10$ fm, which should be caused by the short-range repulsion.  

\begin{figure}[t]
\centering
\includegraphics[width=11.0cm,clip]{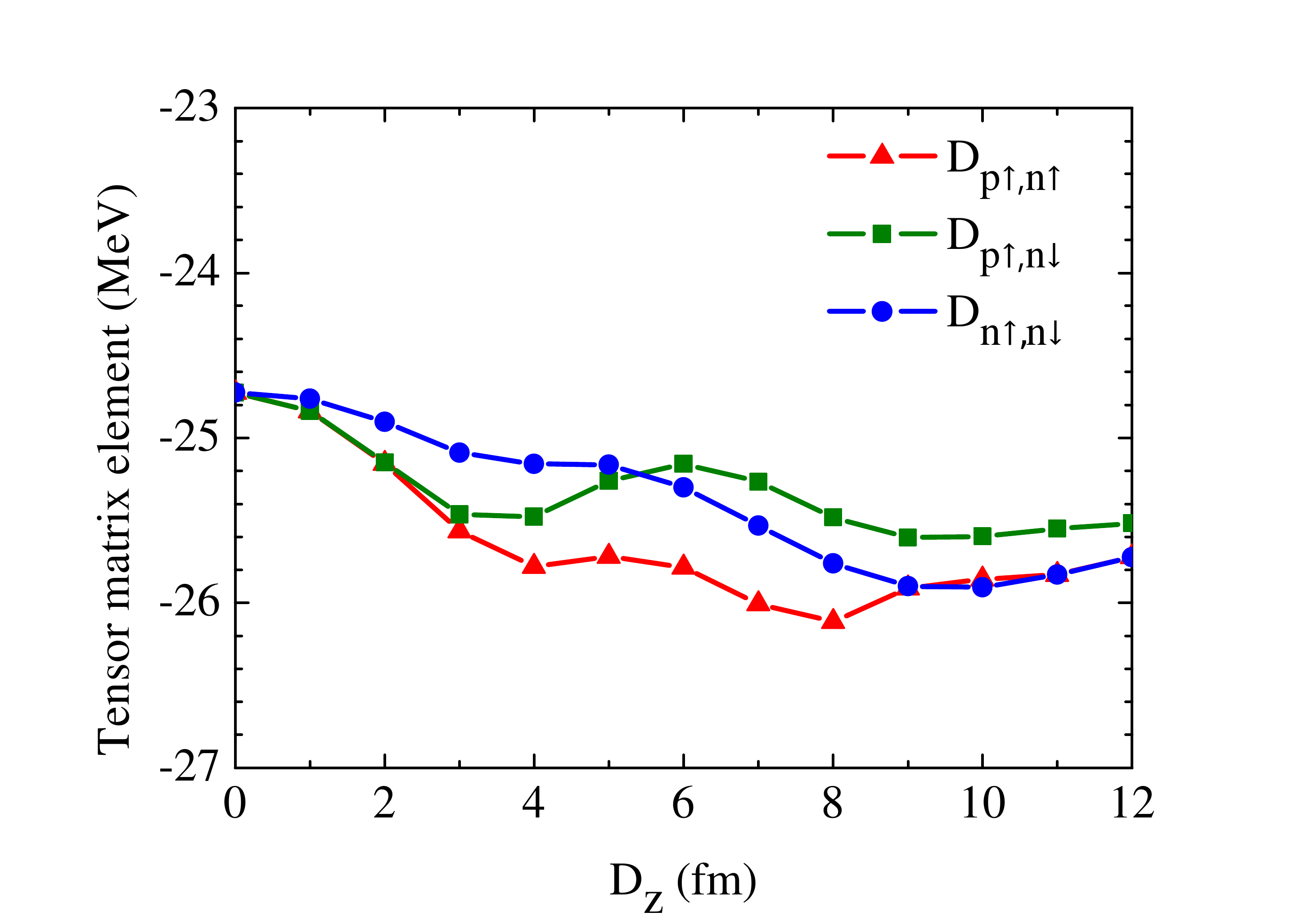}
\caption{Tensor matrix elements of ${^3}$H as functions of the shift $D$. Notations are the same as used in Figure \ref{fig:e-dz}.}
\label{fig:t-dz}
\end{figure}

In Figure \ref{fig:t-dz}, we show the tensor matrix elements as functions of the shift parameter $D_z$, similar to the case of Fig. 2. It is found that the $p$-$n$ pairs commonly show the two minima; one minimum is located at 4 fm, which comes from the tensor correlation in the high-momentum pair. Another minimum around 8-10 fm comes from the short-range correlation in the high-momentum pair, while the correlation functions in the TOAMD part contribute to describe the tensor correlation.

\begin{figure}[t]
\centering
\includegraphics[width=11.0cm,clip]{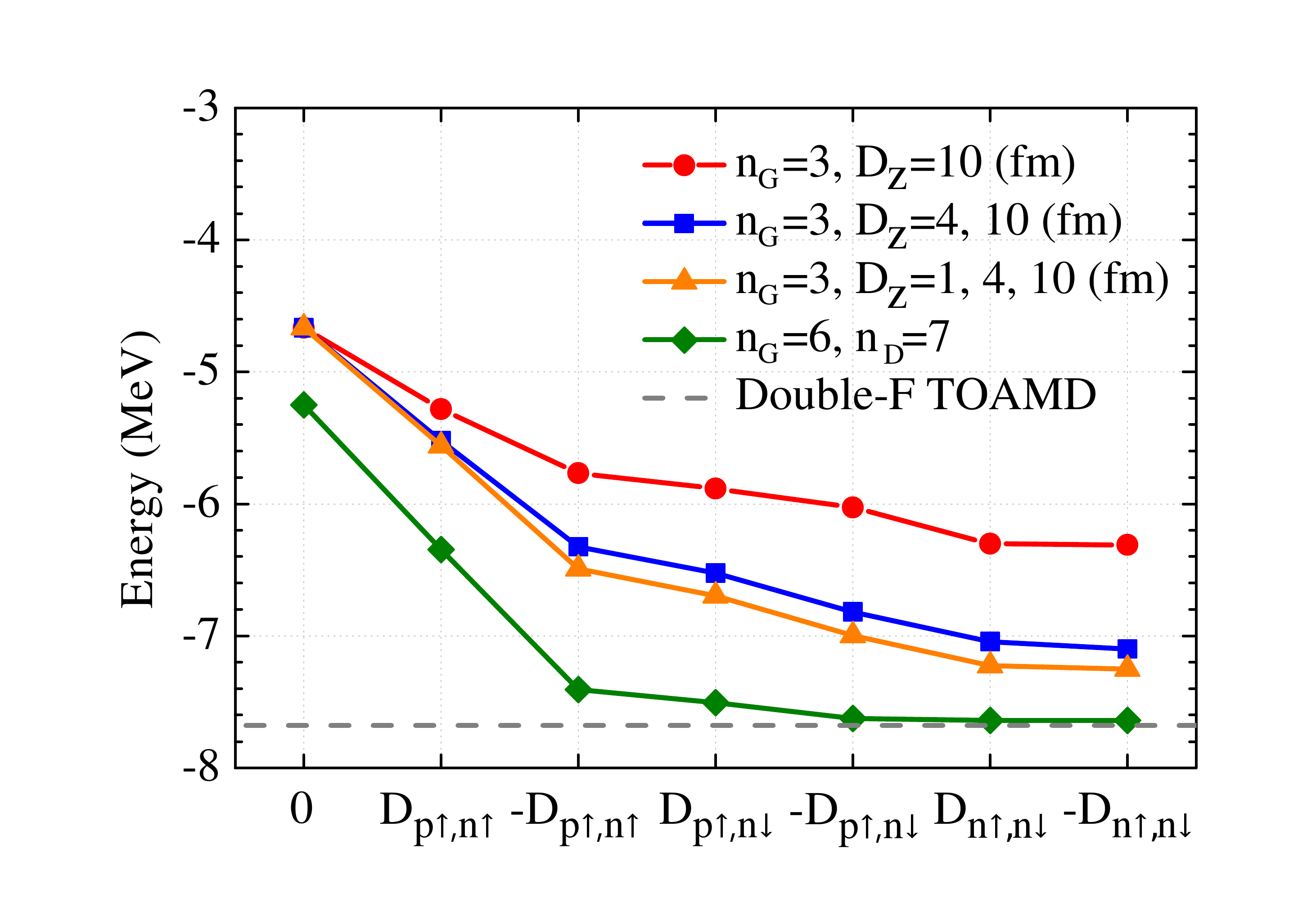}
\caption{Energy of the ${^3}$H nucleus calculated with HM-TOAMD using the bare interaction AV8$^\prime$ with successively adding the various kinds of high-momentum pairs. All the imaginary shifts $\vec D$ are adopted to be parallel with the $z$-axis.  $\pm D$ denote different high-momentum pair shifts in Eq.~(\ref{FH}). Label of each line denotes the values of $|\vec D|$ superposed. We set the range parameter $\nu=0.14$ fm$^{-2}$.}
\label{fig:e-pair}
\end{figure}

To obtain accurate solution of the ${^3}$H wave function, we superpose the basis states having different $\vec D$s with the number of $n_{\rm D}$ for three kinds of pairs in Eq.\,(\ref{FH}). First, we consider only the imaginary shift $\vec D$s along the $z$-axis and show the results by adding successively the different combination of high-momentum pairs in Figure ~\ref{fig:e-pair}.  When we include no high-momentum pairs, which corresponds to the horizontal coordinate 0 in Figure \,\ref{fig:e-pair}, the HM-TOAMD wave function in Eq.\,(\ref{hm-toamd}) reduces to the Single-$F$ TOAMD wave function as $(1+F_S+F_D)|\Psi_\text{AMD}\rangle$ and the energy of ${^3}$H is -4.67 MeV.  After the successive addition of high-momentum pairs into the wave function of HM-TOAMD, the total energy variationally decreases and gets closer to the value given by the Double-$F$ TOAMD. The addition of the first negative shift $-D_{p\uparrow,n\uparrow}$ restores the parity symmetry of the total wave function combined with the positive shift $D$, which contributes more than 1 MeV as shown in the green line with $n_{\rm G}$=6 and $n_{\rm D}$=7. This result shows the importance of the parity symmetry of the HM-AMD basis.  It is also true for rotational symmetry, for which the angular momentum projection could also contribute more than 1 MeV in the final calculation. In Figure \,\ref{fig:e-pair}, it is also shown clearly that the first five kinds of pairs contribute almost all the energy gains compared to the Single-$F$ TOAMD result.  The addition of last pair shift $-D_{n\uparrow,n\downarrow}$ provides tiny improvement of the total energy, which is only 8 keV in the green line. This convergence shows that the first five kinds of pairs are enough to provide accurate descriptions of the NN correlations. It is also found that the energy of $^{3}$H is not improved by adding more high-momentum basis states having $\vec D$s along the $x$- or $y$-axis. Hence, it can be concluded that the model space of high-momentum pairs along the $z$-direction, parallel to the nucleon spin, is already enough in HM-TOAMD. This property is different from the previous analysis of $^4$He with the pure HM-AMD for tensor correlation \cite{myo17e}, which shows the importance of both directions of spin-parallel ($z$) and spin-perpendicular ($x$ or $y$) cases.

In Figure \,\ref{fig:e-pair}, we start our calculation with $n_{\rm D}$=1 using the imaginary shift $D_z=$10 fm {{($k_z=$2.8 fm$^{-1}$)}}, which is favored to express the short-range correlation and corresponds to the red line in the figure. After superposing all the six kinds of high-momentum pairs commonly keeping this shift value, the energy of ${^3}$H is -6.3 MeV, improved by 1.6 MeV compared with the Single-$F$ TOAMD. The further addition of the high-momentum basis state with $D_z=$4 fm {($k_z=$1.12 fm$^{-1}$)}, in which the tensor correlation favored, provides additional improvement by 0.8 MeV for the total energy.  Also, we see contribution of about 0.15 MeV from the $D_z=$1 fm {($k_z=$0.28 fm$^{-1}$)} pairs, this small shift in the momentum space can express the low momentum components of pairs in the ${^3}$H nucleus. After superposition of $n_{\rm D}$=7 with different $D_z$s for all kinds of high-momentum pairs, as shown by the green line in Figure \,\ref{fig:e-pair}, we get the final result for the energy of ${^3}$H  as -7.64 MeV, which is 40 keV higher than the Double-$F$ TOAMD result -7.68 MeV \cite{myo17a} and 120 keV higher than the GFMC result -7.76 MeV \cite{wiringa01}. We list in detail our present results in Table\,\ref{tab:compare} with other calculations, for the Hamiltonian components and root-mean-square of matter radius.  Each component is found to nicely reproduce the corresponding values of Double-$F$ TOAMD and few-body calculation. In particular, the present HM-TOAMD provides the almost equivalent solutions to those of Double-$F$ TOAMD, which indicates the reliability of HM-TOAMD.

\begin{table}[t]
\begin{center}
\caption{Energies of $^3$H (ground state) of HM-TOAMD, Double-$F$ TOAMD ($F^2$-TOAMD) \cite{myo17c} and few-body calculations \cite{wiringa01,suzuki08} in units of MeV.
The radius is in units of fm.}
\label{tab:compare} 
\begin{tabular}{cccc}
\noalign{\hrule height 0.5pt}
        & HM-TOAMD    & $F^2$-TOAMD\cite{myo17a,myo17c}    &Few-body\cite{wiringa01,suzuki08}   \\
\noalign{\hrule height 0.5pt}
Energy  &~$-7.64$~ &~$-7.68$~ &~$-7.76$~\\
Kinetic &~~$47.29$~ &~~$47.21$~ &~~$ 47.57$~\\
Central &~$-22.47$~~ &~$-22.44$~~ &~$-22.49$~~\\
Tensor  &~$-30.60$~~ &~$-30.60$~~ &~$-30.84$~~\\
LS      &~$-1.87$~ &~$-1.86$~ &~$-2.00$~\\ \hline
Radius  &~~~$1.73$~  &~~~$1.75$~ &~~~~$1.75$~~\\
\noalign{\hrule height 0.5pt}
\end{tabular}
\end{center}
\end{table}

\begin{figure}[b]
\centering
\includegraphics[width=11.0cm,clip]{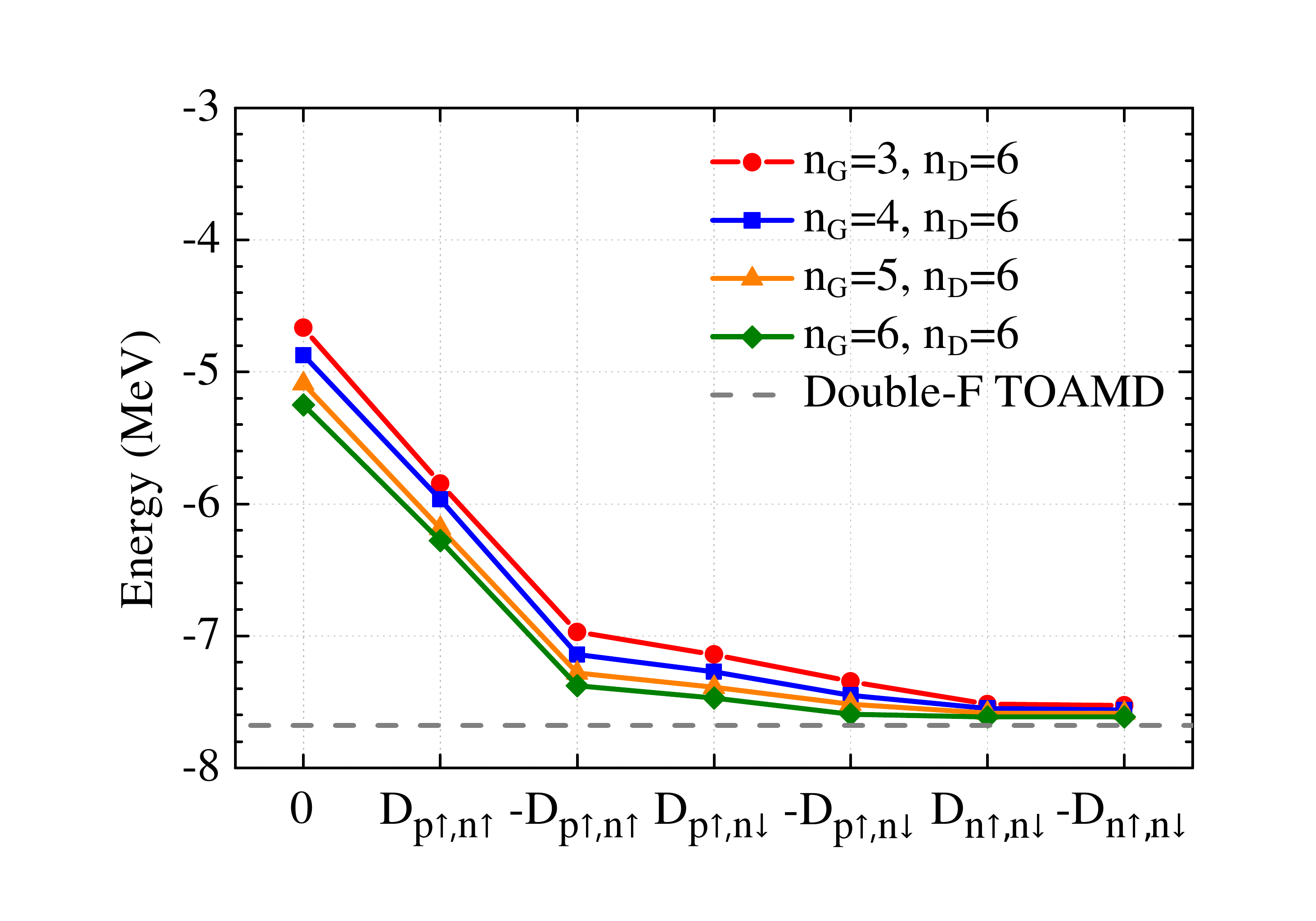}
\caption{Energy of ${^3}$H nucleus calculated with HM-TOAMD using the bare AV8$^\prime$ interaction with respect to successively adding the various kinds of high-momentum pairs.  All the imaginary shifts $\vec D$ are adopted to be parallel with the $z$-axis.  $\pm D$ denote different kinds of the high-momentum pair shifts defined in Eq.~(\ref{FH}). Six shift parameters $D_z$ are superposed for each high-momentum pair. Label of each line denotes the number of Gaussian expansions $n_{\rm G}$ in $F_D$ and $F_S$. We set the range parameter $\nu=0.14$ fm$^{-2}$.}
\label{fig:e-pair-ncr}
\end{figure}

We further discuss the relation between the correlation functions $F_D, F_S$ and the high-momentum pairs.  In the above calculations as shown in Figure  \ref{fig:e-pair}, the correlation functions $F_D$ and $F_S$ are expanded by using the Gaussian functions with the number of $n_{\rm G}=3\sim6$, which is smaller than the choice of $7\sim13$ bases in our previous TOAMD calculations \cite{myo17a,myo17b,myo17c,myo17d}. In Figure \ref{fig:e-pair-ncr}, we show the results with different Gaussian numbers $n_G$ for the correlation functions while keeping the number of high-momentum pair as $n_D=6$. When we include no high-momentum pairs at point "0" in the horizontal axis, we can see clear differences between the energies obtained with different $n_{\rm G}$. After we include all kinds of high-momentum pairs, these differences in the total energy become very small and the resulting energies are almost the same with each other. This result indicates that the small model space of the correlation functions to express the NN correlations, due to small number $n_{\rm G}=3$ in $F_D$ and $F_S$, can be compensated by using high-momentum pairs with larger $n_{\rm D}$. Thus, essentially the $F_D, F_S$ and the high-momentum pairs play a role for describing the same NN correlations and compensate each other. Hence, we can choose flexibly the basis numbers $n_{\rm G}$ for $F_D$ and $F_S$ and $n_{\rm D}$ for high-momentum pairs to set the reasonable model space which can describe the NN correlations of nuclei. In future study of the nuclei of $p$-shell including the clustering states, we can benefit from this flexibility of model space in HM-TOAMD. By utilizing the high-momentum pair, we also simplify the calculations with multiple correlation functions, such as the triple product case and beyond.

\vspace*{0.2cm}
\noindent{\mbox{4.\,\, {\it Conclusion}}}~
In conclusion, we have formulated a new {\it ab initio} framework with the hybridization of tensor-optimized  and high-momentum AMD, namely ``HM-TOAMD". The NN correlations in the wave function is described by using both the correlation functions in the TOAMD approach and the high-momentum pairs in the HM-AMD approach. In the present hybrid description, the energy curves of the ${^3}$H nucleus as functions of high-momentum components show clearly the effects of tensor interaction and strong short-range repulsion. We also find that the convergence of the solutions can be obtained by including only the high-momentum pairs in the $z$-direction, parallel to the spin direction of nucleons. With the AV8$^\prime$ interaction, total energy, Hamiltonian components and matter radius obtained in HM-TOAMD agree very well with other {\it ab initio} frameworks such as the Double-$F$ TOAMD, GFMC and few-body calculations. We also show the flexibility of the model space in this new hybrid method to reduce the total basis number for the numerical calculation.  This is essentially important for the future {\it ab initio} studies of $p$-shell nuclei and nuclei with clustering structures in which the triple and further products of correlation functions must be included.

\section*{Acknowledgments}
$\,\,$This work was supported by the JSPS KAKENHI Grants No. JP15K05091, No. JP15K17662, and No. JP16K05351. One of the authors (M.I.) is supported by the Grants-in-Aid for Young Scientists (B) (No. 15K17671) and Grant-in-Aid for JSPS Research Fellow (No. 16J05297) from the Japan Society for the Promotion of Science. The numerical calculations were performed on the high performance computing server at RCNP, Osaka University.

\nc\PTEP[1]{Prog.\ Theor.\ Exp.\ Phys.,\ \andvol{#1}} 
\nc\PPNP[1]{Prog.\ Part.\ Nucl.\ Phys.,\ \andvol{#1}} 

\end{document}